\documentstyle[multicol,aps,prl]{revtex}
\begin{document}
\tightenlines
\title{Early time behavior of the order parameter coupled to a conserved density:  A study in  a semi-infinite geometry}
\author{Sutapa Mukherji\cite{eml1}}
\address{
Fachbereich Physik, Universit\"at Gesamthochschule Essen, 45117 Essen, Germany
}
\maketitle
\today
\begin{abstract}
We study the short time behavior of the order parameter coupled to a conserved field in semi-infinite geometry. The short time exponent, obtained by solving the one loop differential equations for the conserved density and the order parameter, agrees with the  prediction from a scaling argument based on short distance expansion. The scaling analysis further shows that this  exponent  satisfies a  scaling relation similar to that known in the case  of a nonconserved order parameter without any coupling.
\end{abstract}
\nopagebreak
\widetext
\begin{multicols}{2}
\section{Introduction}
Questions related to   relaxations in  critical dynamics, especially critical slowing down, and the  nature of  transport coefficients have drawn attention due to the unusual properties of a system as its critical point is approached.   So far the long time relaxation   has been the object of primary focus until recent past  when  the short time  relaxation was  found to reveal new universal behavior. This process sets in after the microscopic relaxation processes which are to be described by the microscopic theory. Such a short time property was first observed in numerical simulations \cite{huse} and was  explicitly  calculated for a purely relaxational model with an $n$ component nonconserved  order parameter field $\phi({\bf x},t)$ (model A) \cite{jss,diehlrit}. The short time relaxation involves a new critical exponent which  does not follow from any scaling relation of the known exponents. Being motivated by the study of  boundary critical phenomena \cite{diehlrev}, the original study of the universal short time behavior was based on the consideration of a boundary in the "time" like coordinate.
In the Renormalization Group framework there are  additional singularities  located on the time surface. In general, for the  order parameter one may write the  scaling form \cite{jss}
\begin{equation}
m= \langle \phi({\bf x},t) \rangle
 \sim t^{-\beta/\nu\zeta} f(m_0 t^{x_0/\zeta}),
\end{equation}
where $m_0$ is the initial "magnetization"  with  the scaling dimension $x_0$ and $\zeta$ is the dynamic exponent \footnote{to avoid any confusion with the coordinate $z$, we have denoted the dynamical exponent by $\zeta$ instead of $z$}. $\beta$ and $\nu$ are the usual bulk critical exponents associated with magnetization and length scale respectively \cite{diehlrev}. The function $f(a) \sim a$ for $a \rightarrow 0$ and thus for short time $m\sim m_0 t^{\theta}$ with  $\theta=x_0/\zeta-\beta/\nu\zeta$.
For $a\rightarrow \infty$, $f(a) \sim constant$ and the usual long time relaxation is recovered. For $2<d<4$, since $x_0>\beta/\nu$,  initially the  magnetization  increases upto a certain time $t_0\sim m_0^{-\zeta/x_0}$. In a similar fashion the autocorrelation function $C(t)=\langle \phi({\bf x},t) \phi({\bf x},0) \rangle$ has the short time behavior
$C(t) \sim  t^{\theta-d/\zeta}$ \cite{jss}.

It is well understood now that semi-infinite systems, which extend over  infinite space in $d-1$ dimensions denoted by $r$ and over only the positive half space in the $z$  direction ($z\ge 0$), have critical behavior close to the surface  drastically different from the bulk \cite{diehlrev}. Detail   field theoretical studies show that these differences arise from the need for an additional renormalization factor for  field  to cure the new uv singularities  caused by the surface. Depending on the value of the surface interaction constant,  conventionally denoted by $c$,  there are different  universality classes associated with the surface ordering.  These are named as ordinary  ($c>0$), special ($c=0$) and extraordinary ($c<0$) transitions. In the ordinary transition the surface orders along with the bulk and in  the extraordinary transition the surface orders before the bulk. At the  special point $c=0$,  there is a different set of exponents.  The universal short time behavior is also modified  depending on the surface universality class  considered. As has been shown for model A \cite{uwe,satya}, in the case of the special transition the order at short time grows with time whereas in the ordinary transition it decays with time. By using the short distance expansion (SDE),  introduced by Diehl in the problem of boundary critical phenomena, the  short time exponent  in  semi-infinite model A can be shown \cite{uwe} to satisfy a scaling relation involving the bulk short time exponent and the  static exponents.

In this paper we are primarily concerned with the universal short time behavior of  a  relaxational model with a conservation law. This is model C (as classified by Halperin et al.\cite{halp} ), where the $n$ component order parameter is coupled to a nonordering conserved field.  We shall concentrate on the semi-infinite geometry.  In reality such models  describe  binary alloys undergoing order-disorder transition where  the concentration field (not the order parameter) plays the role of the conserved density \cite{eisen} or systems with mobile conserved impurities or uniaxial Ising antiferromagnets etc. In the static limit, since the conserved field can be integrated out,   model C becomes identical to the static limit of  model A with shifted coupling constants.The static bulk and surface exponents therefore can be simply borrowed from  static limit of model A. In the dynamics of model C,  the coupling of the order parameter with the conserved field plays an important role and depending on the stable fixed points of the parameters,  the $(n,d)$ plane can be separated into different regions where e.g. the relaxation rate, dynamical exponents are different \cite{brez}. The  short time behavior for the bulk  model C  has been studied by  Oerding and Janssen \cite{oerd} who,  using the field theoretic  renormalization group technique, obtain a new universal short time exponent (denoted as $\theta_c$ in the following) for the order parameter relaxation.
The dynamics of the semi-infinite model C \cite{xiong}  has been found to be the same as the bulk dynamics with static exponents same as model A at different universality classes of surface transitions.

The short time dynamics and its universal features  are important in quenching experiments where the system is taken from an unstable ordered state at high temperature to, say,  the critical temperature. In numerical simulations \cite{zheng}, the short time dynamics is relatively easier to observe because in this time regime the critical slowing down does not set in. In several cases the results from early time dynamics gave good estimates for not only the short time but also the bulk exponents. Therefore the knowledge about the  short time exponent for model C and its relation to other critical exponents is expected to be useful for numerical  and experimental work on antiferromagnetic systems \cite{sen}, binary alloys \cite{dosch}, and other systems  where there are  coupling between the order parameter and a conserved density field. 

 A priori it is not  clear what role a conserved density would play  in the short time dynamics of the semi-infinite model C. Another related question is whether the short time dynamics can be explained  by SDE in a similar manner as  model A. By solving  one loop equations of motion for the conserved density  and the order parameter, we show that there are  certain subtleties  in this situation.
Close to the surface the conserved density has spatial variation not coming from SDE. However, this  contributes to the order parameter equation significantly and finally leads to the order parameter relaxation consistent with SDE. Another interesting goal of  studying  model C is to observe the shape of  the conserved density profile. This question  is partially answered in our analysis in a region very close to the surface.

The dynamics of the order parameter field and the conserved nonordering field follows from the hamiltonian
\begin{eqnarray}
{\cal H}[\phi,E]=\int d^d x \big\{\frac {(\nabla \phi)^2}{2}+\frac{\tau}{2} \phi^2+\frac{g}{4!}\phi^4+\frac{E^2}{2}+\frac{\gamma}{2} E \phi^2 \big\}\label{hamb}
\end{eqnarray}
and the Langevin equations
\begin{eqnarray}
\partial_t \phi({\bf x},t)=-\lambda \frac{\delta {\cal H}}{\delta \phi({\bf x},t)}+\zeta({\bf x},t)\\
\partial_t E({\bf x},t)=\lambda\rho\nabla^2(\frac{\delta {\cal H}}{\delta E({\bf x},t)})+\eta({\bf x},t)
\end{eqnarray}
The Gaussian random noise $\eta(x,t)$ and $\zeta(x,t)$ have zero mean and  correlations
\begin{eqnarray}
\langle \zeta({\bf x},t) \zeta({\bf x}',t')\rangle=2 \lambda \ \delta({\bf x}-{\bf x}')\delta(t-t')\\
\langle \eta({\bf x},t) \eta({\bf x}',t')\rangle=-2\lambda\  \rho \ \nabla^2\delta({\bf x}-{\bf x}')\delta(t-t').
\end{eqnarray}

The generating functional in terms of the response fields $\tilde \phi(x,t)$ and $\tilde E({\bf x},t)$ is \cite{martin,domin}
\begin{eqnarray}
&&{\cal J}[\phi, \tilde \phi,m,\tilde m]=\int dt \int d^d x[\tilde \phi\{\partial_t \phi({\bf x},t)+\lambda(\tau-\nabla^2) \phi+\nonumber\\
&&\frac{\lambda g}{3!} \phi^3+\lambda \gamma \phi({\bf x},t) E({\bf x},t)\}-\lambda \tilde \phi^2({\bf x},t)+\tilde E({\bf x},t) \partial_t E({\bf x},t)-\nonumber\\
&&\lambda \rho(\nabla^2 \tilde E({\bf x},t))(\frac{\gamma}{2} \phi({\bf x},t)^2+E({\bf x},t))-\lambda \rho (\nabla \tilde E)^2]\label{gener}
\end{eqnarray}
 The correlation propagator in the bulk case contains an equilibrium part which is translationally invariant  in time and a non equilibrium mirror symmetric part. To be brief, we  mention only the correlation propagator for the field $\phi$ and refer the reader to \cite{oerd} for the other response and correlation propagators. In Fourier space the correlation propagator is
\begin{eqnarray}
C_{\phi}({\bf q},t-t')=\langle \phi({\bf q},t)\phi(-{\bf q},t')\rangle=
[e^{-\lambda(\tau+ q^2)\mid t-t'\mid}-\nonumber\\
e^{-\lambda(\tau+q^2)(t+t')}]/(\tau+q^2)
\end{eqnarray}
This propagator corresponds to the Dirichlet initial  condition.

The static version is equivalent to the static limit of model A with $\phi^4$ coupling $\tilde u=g-3\gamma^2$.
We briefly recall,  from the renormalization group analysis of the dynamics of model C \cite{brez},  the results useful for the present work. The dimensionless coupling constants $u$ and $v$  are defined as
$
K_d \tilde u =\mu^{\epsilon} u$ and
$K_d \gamma^2=\mu^{\epsilon} v$,
where $\mu$ is an arbitrary momentum scale and $\epsilon=4-d$ and $K_d^{-1}=2^{d-1} \pi^{d/2} \Gamma[d/2]$.
Broadly there are two distinct regimes in the $(n,d)$ plane depending upon the sign of the specific heat exponent $\alpha$. For $\alpha<0$ the stable fixed points are $u^*=6 \epsilon/(n+8)$  and $v^*=0$. Clearly in this regime large scale properties of model C are  same as model A and the dynamic exponent $\zeta=2+O(\epsilon^2)$. For  $\alpha > 0$ there is a stable nonzero fixed point $v^*=2 \epsilon(4-n)/[n(n+8)]$.These two regimes can further be separated depending on the fixed point of $\rho$. For the details on these aspects we refer the reader to \cite{brez}.  For one component order parameter $\rho^*=1+O(\epsilon)$ and the conventional scaling $\zeta=2+\alpha/\nu$ holds good.

In an attempt to understand  the short time exponent for model C in a semi-infinite geometry using a scaling argument based on a SDE, we write the  scaling form for the magnetization
\begin{eqnarray}
m(z,t,m_0) \sim t^{-\beta/\nu \zeta} {\cal F}(z/t^{1/\zeta},m_0 t^{x_0/\zeta}).\label{scalfn}
\end{eqnarray}
Since the conserved density profile has a nontrivial behavior solely due to the coupling with the order parameter field, we have ignored the explicit dependence of the scaling function on the conserved field. Our  approach of solving one loop renormalized equation of motion provides a more rigorous justification for this. Furthermore because of  the scaling of the magnetization close to the surface as  $m \sim t^{-\beta_1/\nu \zeta}$, where $\beta_1$ is the surface magnetization exponent \cite{diehlrev}, we expect ${\cal F}(x,y) \sim x^{a} {\cal F}_1(y)$ such that $a=(\beta_1-\beta)/\nu$. This implies that near the boundary
\begin{eqnarray}
m(z,t,m_0)= z^{(\beta_1-\beta)/\nu} t^{-\beta_1/\nu\zeta} {\cal F}_1(m_0 t^{x_0/\zeta}).
\end{eqnarray}
This result can also be understood by SDE which should hold good for fields with different scaling dimensions on or off the surface.
Since at short time $m$ should be proportional to $m_0$, the short time behavior close to the surface is
\begin{eqnarray}
m(z,t) \sim m_0 z^{(\beta_1-\beta)/\nu} t^{\theta_c+(\beta-\beta_1)/\nu\zeta}.\label{relate}
\end{eqnarray}
 Eqn. (\ref{relate}) shows that  the early time behavior close to the surface is described by  the exponent $\theta_{c1}$ which satisfies the scaling relation 
\begin{equation}
\theta_{c1}=\theta_c+(\beta-\beta_1)/\nu \zeta, \label{scalrel}
\end{equation}
 involving the bulk short time exponent and static exponents.

The above scaling relation is supported by an explicit calculation starting from  the  linearized equations for the order parameter and the conserved density. This approach is elaborated in section II. Singularities appearing at the one loop level  are  taken care of by the renormalization of  appropriate parameters. Solving the  renormalized equations, the short time exponents are then obtained. The details in Section II is important  to appreciate the crucial role played by the conserved density. The justification of the scaling argument and the agreement between the two approaches are  discussed in section III.

\section{ONE LOOP Renormalized equations}
Starting from initial conditions and the  translational invariance in $d-1$ directions, the
one loop equations for  time variations of the averaged  conserved density ${\cal E}(z,t)=\langle E({\bf x},t)\rangle$ and  order parameter  $m(z,t)=\langle \phi({\bf x},t)\rangle$   can be written as
\begin{eqnarray}
&&\frac{1}{\lambda \rho} \partial_t {\cal E}=\partial_z^2 {\cal E}(z,t)+\frac{\gamma}{2} \partial_z^2 \langle \phi^2 (x,t)\rangle \label{energy}\\
&&\frac{1}{\lambda}\partial_t m(z,t)+[\tau_0 m-\nabla^2 m+\frac{u_0 (n+2)}{6} C(z,t) m+\nonumber\\
&&\gamma_0 \langle E(z,t) \phi(z,t)\rangle]=0, \label{order}
\end{eqnarray}
The one loop term introduces spatial variation in the conserved density profile which is flat otherwise.  The one loop contribution $C(z,t)=\langle \phi^2({\bf x},t)\rangle$ consists of bulk and surface parts  as $C(z,t)=C_b(0,t)\pm C_b(2z,t)$ \cite{dd1}, where
\begin{equation}
C_b(z,t)=\frac{1}{(2\pi)^d} \int \frac{d^d q}{\tau_0+q^2}[1-e^{-2\lambda(q^2+\tau)t}] \exp[i k z]\label{bcorr}
\end{equation} and $\pm$ refer to special and ordinary transitions respectively.
The first term on the RHS of (\ref{bcorr}) in the bulk contribution $C_b(0,t)$ has a divergence which has to be absorbed by the renormalization of the temperature. Evaluating the integrals we obtain
\begin{equation}
C(z,t)=\frac{1}{32 \pi^2}[-\frac{1}{\tilde t}\pm \frac{2}{z^2} \exp(-z^2/2\tilde t)], \label{oneloop}
\end{equation}
 where $\tilde t=\lambda t$. Substituting this in (\ref{energy}), we find that  the solution of the conserved density profile is of the form  ${\cal E}(z,t)=\frac{1}{\overline t} F(z^2/{\overline t})$, where ${\overline t}=\lambda \rho t$. Very close to the boundary such that $z\ll t^{1/2}$, $F(x) \sim \mp \frac{\gamma}{32 \pi^2 x}$, where $-$ refers to the special transition. Therefore very close to the boundary we have
\begin{equation}
{\cal E}(z,t)\sim \mp {\gamma}/{32 \pi^2 z^2}\label{thirt},
\end{equation}
for special and ordinary transition respectively. Eqn. (\ref{thirt}) shows the importance of the one loop term from $\phi$ (the inhomogeneous term in  (\ref{energy})) in the behavior of $\langle E \rangle$ near the surface. Since the above  result is  restricted to the regime $z\ll t^{1/2}$, initial condition cannot be reached from this.  At the fixed point of our interest this power law form  associated with a prefactor  $O(\epsilon^{1/2})$ contributes at  $O(\epsilon)$ in the equation for the order parameter.
Though a further analysis about the shape of the conserved density profile away from the surface deserves attention \cite{diehljan}, we here restrict ourselves very close to the surface.

Next we consider the term $A(x,t)=\langle E({\bf x},t) \phi({\bf x},t)\rangle$ which needs to be expanded in order to take into account the  other  $O(\epsilon)$ terms in (\ref{order}). Using the generating functional in (\ref{gener}), we have
\begin{eqnarray}
&&A(x,t)=\langle E({\bf x},t) \phi({\bf x},t)\rangle-
\lambda\gamma\int d^dx_1 \int_0^t dt_1 \nonumber\\ &&\{\langle E({\bf x},t) E({\bf x}_1,t_1)\rangle \langle \phi({\bf x},t) \tilde \phi ({\bf x}_1,t_1)\rangle\langle \phi({\bf x}_1,t_1)\rangle-\nonumber\\
&& \rho   \langle E({\bf x},t) \nabla^2 \tilde E({\bf x}_1,t_1)\rangle \langle \phi({\bf x},t) \phi({\bf x}_1,t_1) \rangle \langle \phi({\bf x}_1,t_1)\rangle\}
\end{eqnarray}
 For convenience we denote the  two terms in the curly bracket by $A_1(x,t)$ and $A_2(x,t)$ respectively. It is apparent that there is no straight forward way to evaluate the last two terms due to their coupled structure. To obtain the contributions of these term in the bulk case, we assume  $\langle \phi({\bf x}_1,t_1)\rangle$  to be space time independent.  This is also justified if, say, the  time dependence has an  universal exponent $\sim t^p$ where $p \sim O(\epsilon)$. In that case from the expansion  $ \langle \phi({\bf x}_1,t_1)\rangle \sim 1+\epsilon \ln t$ it is clear that upto this order of calculation only the time independent piece is the important one.
Now the contributions from the above terms in the bulk case are
\begin{eqnarray}
A_1(x,t)
=-\frac{s_d}{(2\pi)^d}\frac{\gamma}{\tilde t}[ \frac{1}{2(1+\rho)^2}+\frac{1}{4 \rho(1+\rho)}], \nonumber\\
A_2(x,t)
=-\frac{s_d}{(2\pi)^d}\frac{\gamma}{\tilde t}[\frac{\rho}{4(1+\rho)}+ \frac{\rho}{2(1+\rho)^2}],
\end{eqnarray} where $s_d=\frac{2 \pi^{d/2}}{\Gamma[d/2]}$. These two terms correspond to second and third diagrams in Fig 1 of \cite{oerd}. Correspondingly the second terms in the square brackets can be compared with  $O(1)$ parts in the fourth and third terms in Eqn. (32) of \cite{oerd}. These are the contributions from the `initial' part of the correlators.

In the semi-infinite case we have
\begin{eqnarray}
 A_1(z,t)=\pm \frac{  s_d\gamma}{(2\pi)^d }[\frac{2\rho\exp[-z^2/(1 +\rho)\tilde t]}{ 2 z^2(\rho^2-1)}+  \frac{\exp[-z^2/2\rho \tilde t]}{2 z^2(1-\rho)}]\nonumber\\
A_2(z,t)=\pm \frac{ s_d\rho \gamma}{(2\pi)^d }[\frac{2\exp[-z^2/(1 +\rho)\tilde t]}{2 z^2(1-\rho^2)}
 -\frac{\exp[-z^2/2\tilde t]}{2 z^2(1-\rho)}]\nonumber
\end{eqnarray}
Adding all the  bulk and surface contributions  from above we have
\begin{eqnarray}
\frac{1}{\lambda} \partial_t m-\partial_z^2 m+\frac{u}{\ (64 \pi^2)}[-\frac{1}{\tilde t}\pm \frac{2}{z^2}]m+\frac{\gamma^2}{16 \pi^2 \tilde t} m\nonumber\\ \mp \frac{3\gamma^2}{32 \pi^2} \frac{m}{z^2}=0
\end{eqnarray}
Note that for $\gamma=0$, we get back the equation for model A.
This equation  can be solved by assuming a  scaling form
$m(z,t)=U(\tilde t) V(z/t^{1/\zeta})$. Here  we are restricted to  $\zeta=2$. In the bulk limit $z\rightarrow \infty$ we obtain $U(\tilde t) \sim {\tilde t}^{\theta_c}$, where $\theta_c= O(\epsilon^2)$. This agrees with the bulk short time exponent in \cite{oerd}. For $z\ll t^{1/2}$, we find $V(z/t^{1/2}) \sim (z/t^{1/2})^{1-\epsilon/6}$  for the ordinary transition and $V(z/t^{1/2}) \sim (z/t^{1/2})^{- \epsilon/6}$ for the special transition. Thus for magnetization upto $O(\epsilon)$, we have
\begin{eqnarray}
 m(z,t) && \sim m_0 z^{1-\epsilon/6 } \tilde t^{-1/2+\epsilon/12} \ \ {\rm \ for \ ordinary\  transition}\label{ordtime}\\
&&  \sim m_0 z^{-\epsilon/6} t^{\epsilon/12}\ \ {\rm \ for \ special \ transition.}\label{sptime}
\end{eqnarray}

\section{CONFORMITY WITH SCALING ANALYSIS}

A few points in the above perturbative calculation and in the previous scaling analysis need to be re-emphasized here. From the nature of the one loop equation for the conserved density and its solution, it is clear that we are in a time regime where the conserved density profile is controlled by the $\phi$ correlations. As a consequence, the  inhomogeneous equation (\ref{energy}) provides a spatially dependent one loop correction to the conserved density profile ${\cal E}$  with a prefactor of $O(\epsilon^{1/2})$. The conserved field is, therefore,  redundant in the scaling function which involves only scaled variables with $\epsilon$ dependence in various exponents.

We see that the short time exponents in (\ref{ordtime}) and (\ref{sptime}), obtained by solving the differential equations,  are  in agreement with the  prediction from scaling analysis upto the factor $\zeta$ (recall that for bulk model C, $\theta_c=O(\epsilon^2)$ \cite{oerd} and the  static exponents $\beta$ and $\beta_1$ are same as obtained from the static limit of model A). 
Clearly while solving  one loop equations, we are restricted to $\zeta=2$ because of the very nature of the equation. The fact that  $\zeta=2+\epsilon/3$  should appear from the appropriate  propagator renormalization which has not been performed in this simple approach.  Since the static exponent of $z$, obtained from the scaling analysis,  is in clear agreement with the calculation, we use $\zeta=2+\epsilon/3$ in
$V(z/t^{1/\zeta}) \sim (z/t^{1/\zeta})^{- \epsilon/6},  (z/t^{1/\zeta})^{1-\epsilon/6}$ for special and ordinary transitions respectively. Thus we conclude that the short time exponents in the semi-infinite model C are
\begin{eqnarray}
\theta_{c1}&&=\epsilon/6\zeta\ \ \ {\rm for\  special\ transition} \label{sp1}\\
&&= -(1-\epsilon/6)/\zeta\ \ \ {\rm for \ ordinary\ transition}.\label{ord1}
\end{eqnarray}

The characteristics of model C is reflected in the exponent of time, whereas the exponent of $z$ remains same as model A. This is  due to the fact that the  static limits  of model C and  model A are same. We do not discuss more about the behavior of the autocorrelation function. The short distance, short time behavior of this quantity for model A has been explained in \cite{uwe}. Our analysis shows that the  same scaling relation is valid here.

\section{CONCLUSION}
We have obtained the nature of the conserved density profile close to the surface by solving a one loop differential equation. This one loop correction to the flat profile is of $O(\epsilon^{1/2})$. This form of the profile has significant impact on the relaxation of the order parameter in the semi-infinite geometry.  The universal short time exponent $\theta_{c1}$ in a semi-infinite geometry is given in (\ref{sp1}) and (\ref{ord1}).  As in model A the short time exponents for special  and ordinary  transitions differ drastically and they are consistent with the scaling analysis based on short distance expansion.

I thank H. W. Diehl,  S. M. Bhattacharjee and U. Ritschel for many fruitful discussions and suggestions. Financial support from Deutsche Forschungsgemeinschaft (DFG) through Sonderforschungsbereich 237 is acknowledged.

\end{multicols}
\end{document}